\documentclass[CRPHYS,Unicode,manuscript]{cedram}

\title{What is measured when measuring a thermoelectric coefficient?}
\alttitle{Que mesure-t-on en mesurant un coefficient thermoélectrique?}
\author{\firstname{Kamran} \lastname{Behnia}}
\address{Laboratoire de Physique et d'Etude de Mat\'eriaux (CNRS-Sorbonne University)\\
ESPCI Paris, PSL University\\
10 Rue Vauquelin, 75005 Paris, France}
\email[]{kamran.behnia@espci.fr}


\begin{abstract} 
A thermal gradient generates an electric field in any solid hosting mobile electrons. In presence of a finite magnetic field (or Berry curvature) this electric field has a transverse component. These are known as Seebeck and Nernst coefficients. As Callen argued, back in 1948, the Seebeck effect quantifies the entropy carried by a flow of charged particles in absence of thermal gradient. Similarly, the Nernst conductivity, $\alpha_{xy}$, quantifies the entropy carried by a flow of magnetic flux in absence of thermal gradient. The present paper summarizes a picture in which the rough amplitude of the thermoelectric response is given by fundamental units and material-dependent length scales. Therefore, knowledge of material-dependent length scales allows predicting the amplitude of the signal measured by experiments. Specifically, the Nernst conductivity scales with the square of the mean-free-path in metals. Its anomalous component in magnets scales with the square of the fictitious magnetic length. Ephemeral Cooper pairs in the normal state of a superconductor generate a signal, which scales with the square of the superconducting coherence length and smoothly evolves to the signal produced by mobile vortices below the critical temperature. 

\end{abstract}

\begin{document}

\maketitle

\section{Introduction}
Thermoelectricity \cite{MacDonald,Goldsmid2009,Behnia2015b} was discovered in 1821 by Seebeck, who connected a bismuth wire to an antimony wire and found that heating the junction generates a voltage difference between the two free ends of the pair. Thirteen years later, Peltier injected an electric current across a junction of two metals, and detected a temperature difference. In 1854, Thomson (the future lord Kelvin) showed that Seebeck and Peltier effects were manifestations of the same phenomenon. 

The Seebeck coefficient is the ratio of the voltage difference to the temperature difference. The Peltier coefficient is the ratio of the heat current density to the electric charge density. For each material, they can be measured independently at a given temperature. What is now known as the Kelvin relation, states that the Peltier coefficient is equal to the Seebeck coefficient times the absolute temperature. 

Thermoelectricity is fundamentally unavoidable because of the combination of the conservation of energy and the conservation of particle number \cite{Behnia2015b}. Sustaining a temperature difference in presence of mobile particles in equilibrium requires a force impeding the flow caused by an excess of kinetic energy. The potential energy provided by an electric field counters this flow and warrants equilibrium. 

The first remarkable fact about this effect is its small magnitude.  Kelvins are exchanged for mere micro-volts. The Seebeck coefficient has for natural unit the ratio of the Boltzmann constant to the charge of electron: $k_B/e= 86 \mu V/K$. In most metals, the room-temperature Seebeck coefficient is much smaller than this. In semiconductors, on the other hand,  the Seebeck coefficient can be several times $k_B/e$. A second remarkable fact is the absence of any geometric factor in Seebeck and Peltier coefficients. Since they are defined as the ratio of two gradients or two flow densities, they are intensive quantities, which do not depend on the size of the sample.  A third feature is the remarkably delicate evolution of the Seebeck coefficient  with temperature, which remains undeciphered by the state-of-the-art theoretical calculations in most solids. This is remarkable, since such calculations are quite successful nowadays in providing a quantitative description of the ground state of complex crystals. One reason is that in many cases, the thermoelectric response is not purely diffusive. When heat is mostly carried by phonons, the thermoelectric response is contaminated by 'phonon drag' and its rigorous quantification requires a comprehensive grasp of their momentum exchange with mobile electrons \cite{Herring}.

The aim of the present paper is to give an account of the order of magnitude of the thermoelectric response in several different contexts. During the last two decades, numerous experiments have made it clear that the magnitudes of the longitudinal (Seebeck) and transverse (Nernst) thermoelectric response are set by a combination of fundamental constants and material-dependent length scales. This way of approaching the thermoelectric response follows Landauer’s picture of transport as a transmission phenomenon \cite{imry}, which has transformed the way we understand the electrical (and the thermal) conductivity in one, two, and three dimensions.

\section {The thermodynamic origin of a transport coefficient}
In mid-twentieth century, Herbert Callen, applying Onsager's reciprocal relations \cite{Onsager1,Onsager2,CASIMIR} to thermoelectricity, proposed a `simple intuitive interpretation’ of the Seebeck coefficient. It is the ratio of entropy flow to particle flow in the absence of a thermal gradient \cite{Callen1948,Callen1952}. 

At a first sight, the  ostentatious `absence of thermal gradient' in Callen's definition may look puzzling. To measure the Seebeck coefficient, the experimentalist needs a temperature gradient.  But thermoelectricity is only a correction  to heat propagation in solids, and often a minor one. The thermal gradient generates a heat flow (or vice versa, depending on the way you see the causality between the two) and this heat flow does not require particle flow. The Seebeck coefficient quantifies the amount of entropy which particles can carry with no help from the thermal gradient. The Onsager picture of reciprocity puts the Kelvin relation , between Peltier and Seebeck coefficients, on firm grounds. Now, if the Peltier coefficient is the thermal energy density flow (in W.m$^{-2}$) divided by the charge density flow (in A.m$^{-2}$), dividing it by temperature yields the Seebeck coefficient, i.e. the ratio of the entropy density flow (in W.K$^{-1}$.m$^{-2}$) to the charge density flow. 

Electric or thermal conductivity cannot change sign. During irreversible processes in a solid, energy can be dissipated, but not created. In contrast, the Seebeck and the Peltier coefficient can be either positive or negative. The flow of energy can be transported by charge carriers of either sign:  Positive when the response is electron-like, or negative when it is hole-like. In the first case, the overall response is dominated by occupied electronic states and in the second by the unoccupied ones.

Defining the Seebeck coefficient as a measure of entropy carried by a flowing charge carrier raises a question. What about the scattering time (or the mean-free-path)? Aren't we in presence of a transport coefficient? The answer is that the entropy in question includes the information contained in the scattering suffered by the mobile particle. 

Let us consider a very simple case, which is the `wrong' sign of the Seebeck coefficient in noble (and most alkali) metals  \cite{Robinson1967,Xu2014,Behnia2015b,Xu2020}. Electrons in lithium, copper, silver and gold are very close to a gas of free electrons. Yet the experimentally-measured Seebeck coefficient of these metals is positive like a gas of hole-like particles. How come?

The answer to this question requires an examination of the Fermi-Dirac distribution of electrons (Fig. 1). Quantum mechanics condemns a fermion to avoid occupying a state occupied by another fermion. Therefore, electrons occupy all available states up to what is called the Fermi level, above which the empty states begin. Such a distribution of states works for both electrons in metals and for neutrons in a neutron star. The step-like transition separating the occupied to unoccupied states is broadened by warming. Thus, applying a temperature difference along a rod of copper generates a broader Fermi-Dirac distribution on the hot side and a narrower one on the cold side. An electric field is required to counter this imbalance. But what should be its sign? 

The largest entropy in a Fermi-Dirac distribution belongs to the state at the middle of the step-like transition whose probability of occupation is close to 1/2. \footnote{The von Neuman entropy of a Fermi-Dirac distribution $f(\epsilon)$ ($S_{VN}= k_B f(\epsilon)lnf(\epsilon)$) peaks when $\epsilon=\mu$. }  For those at higher or lower probability, the chance of flipping towards occupation or vacancy is equal. Thus, the state with highest amount of entropy, which remains identical on both sides, does not contribute to the thermoelectric response. On the warm side, the occupation probability increases for high-energy states and decreases for the low-energy states. The reverse happens on the cold side. Since the entropy content of the states decreases symmetrically with their distance from the middle of the step, the entropy imbalance is insignificant. The fundamental reason of the weakness of the thermoelectric response in metals is the way the Fermi-Dirac distribution concentrates its entropy in the vicinity of the chemical potential. 

 \begin{figure}[tbp]
 \includegraphics[width=0.9\linewidth]{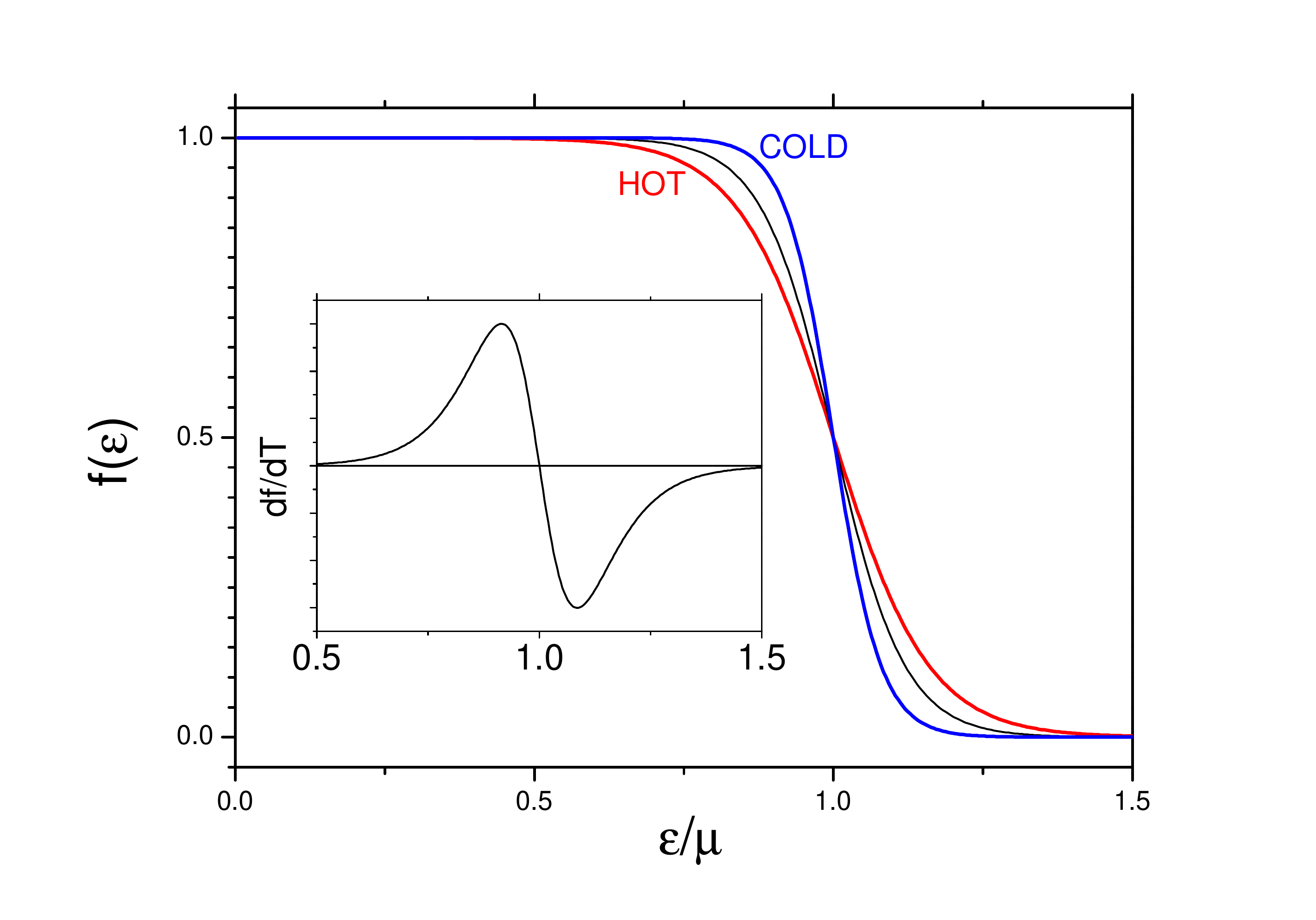}
 \caption{The Fermi-Dirac distribution broadens with heating. The inset shows its temperature derivative. The change in occupation probability with opposite signs below and above the chemical potential is symmetrical.}
 \label{fig:FD}
 \end{figure}
 
The sign of the Seebeck coefficient is set by the fact that states below and above the Fermi level do not have the same density, velocity and/or mean-free-path.  The 'wrong' sign seen in noble metals arises because the mean-free-path of electrons in these solids has a sizeable energy dependence \cite{Robinson1967,Xu2000}. The mean-free-path  decreases fast enough to compensate for increasing velocity and density.  

Such a situation, which leads to opposite signs for the Hall and the Seebeck coefficient, is encountered in numerous occasions in metallic solids. On the other hand, in semiconductors (where the electron gas is not dense enough for the Pauli exclusion principle to hold), the sign of carriers yielded by the Hall and the Seebeck coefficient match each other. Note that here $T \gg T_F$. Indeed, in contrast to metals, intrinsic semiconductors host charge carriers only at a finite temperature and this temperature exceeds the degeneracy temperature.

Coming back to Callen's ‘simple intuitive interpretation’, what we just saw in the case of noble metals, is that the entropy of an electron near the Fermi energy can contain information regarding the energy-dependence of its scattering time near the Fermi energy. This reverses the sign of the Seebeck coefficient, but does not modify its order of magnitude. A more spectacular case is the Kondo effect \cite{Hewson}, where the scattering of electrons by magnetic impurities leads to an order of magnitude amplification of the thermoelectric response. The amplitude of the Seebeck coefficient in gold  is sensitive to the presence of iron impurities with a concentration as low as  0.001 ppm \cite{Kopp_1975}. There is a large difference between the amount of information in the vicinity of the Fermi level between an electron traveling in pure gold and the one in gold with a tiny concentration of iron impurities. The information is generated by many-body entanglement between the spins of the impurity and the electrons in the Fermi sea.
\section{The paradigm shift instigated by Landauer}
In  1996,  reviewing the advent of quantum point contacts, Henk van Houten and Carlo Beenakker gave the following account of how this was accompanied by a change in theoretical understanding of electric conduction:

\textit{On the theoretical side, there was the debate whether a wire without impurities could have any resistance at
all. Ultimately, the question was: ``What is measured when you measure a resistance?'' The conventional point
of view (held in the classical Drude-Sommerfeld or the quantum mechanical Kubo theories) is that conduction
is the flow of current in response to an electric field. An alternative point of view was put forward in 1957 by Rolf
Landauer (IBM, Yorktown Heights), who proposed that “conduction is transmission”.} \cite{vanhouten}

Landauer's conceived conductance of a sample as emanating from its transmissive behavior \cite{Landauer1992}. This  led to his formula for conductance, $G$, which includes the quantum of conductance: 
\begin{equation}
 G= T N_C \frac{2e^2}{h} 
\end{equation} 

Here, $e$ and $h$ are the elementary charge and the Planck constant. The ratio $\frac{e^2}{h}$, which has the units of conductance (Siemens or inverse of Ohm) sets a quantum scale for the measured conductance. $T$ is the transmission probability, which can take any value from 0 to 1. N$_C$ is an integer which represents the number of conduction channels. The quantization of conductance, explicit in this approach when $T=1$, has been observed in experiments on point contacts \cite{vanhouten} and break junctions \cite{Ohnishi}. 

This approach to conductance illuminates the Boltzmann-Drude expression for electrical conductivity, $\sigma$ of a metal \cite{Ashcroft76}:
\begin{equation}
 \sigma = \frac{ne^2\tau}{m^*}  
 \label{Drude}
\end{equation} 
Here, $n$ represents the density of charge carriers, $m^*$ is their mass and $\tau$ their scattering time. The natural scale of conductance, namely $\frac{e^2}{h}$, does not appear in this expression. However, it is easy to show that Eq. \ref{Drude} is strictly equivalent to :
\begin{equation}
 \sigma = A_d\frac{e^2}{h}k_F^{d-1}\ell
 \label{Land}
\end{equation} 
Here $d=1,2,3$ represents the dimensionality of the system. $k_F$ is the Fermi wave-vector, $\ell$ is the mean-free-path and A$_d$ is a dimension-dependent numerical factor of the order of unity. Besides explicitly including the natural units of conductivity, Eq.\ref{Land} has other advantages over Eq.\ref{Drude}. It implies that conductivity does not scale with the number of electrons inside the Fermi \textit{sea} (which is $\propto n$), but with the number of the electrons at the Fermi \textit{surface} (which is $\propto k_F^{d-1}$). It does not give the wrong impression that metals with heavy electrons \cite{Hewson,Flouquet2005OnTH} have a lower conductivity than those hosting lighter electrons. Finally, one can see that even the diffusive conduction of charge is about the transmission of a quantum of conductance. In one dimension, this is represented by The product $\ell$, the distance an electron travels without colliding. In two dimensions, the number of channels scales with $k_F$ and in three dimensions, with $k_F^2$.
\section{The Seebeck coefficient }
What about the thermoelectric response ? The common expression for the Seebeck coefficient of a degenerate electron gas is \cite{Ashcroft76}: 
\begin{equation}
S _{(k_BT \ll E_F)}=\pm \frac{\pi^2}{3}\frac{k_B}{e}\frac{k_B T}{E_F}
 \label{S1}
\end{equation} 
This equation tells us that in a Fermi-Dirac distribution, available  entropy is restricted to a thermal window in the vicinity of the Fermi energy. Now, using the expressions for the Fermi wavelength, $\lambda_F=\sqrt{\frac{h^2}{2m^*E_F}}$ and the de Broglie thermal length, $\Lambda=\sqrt{\frac{h^2}{2\pi m^*k_BT}}$, one can show that Eq.\ref{S1} is  strictly equivalent to this one \cite{Behnia2015b}:
\begin{equation}
S_{(\Lambda \gg \lambda_F)}= \pm A_S \frac{k_B}{e}\frac{\lambda_F^2}{\Lambda^2}
 \label{S2}
\end{equation} 
$A_S$ is a numerical constant. For a three-dimensional free gas (of electrons or holes), it is  equal to $\frac{\pi}{3}$. As seen above when discussing the Seebeck coefficient in noble metals, the sign and the precise amplitude of this prefactor changes according to the energy dependence of the relaxation time (and the density of states) near the Fermi level.  The order of magnitude, however, does not.

\begin{figure}[tbp]
 \centering
 \includegraphics[width=0.75\linewidth]{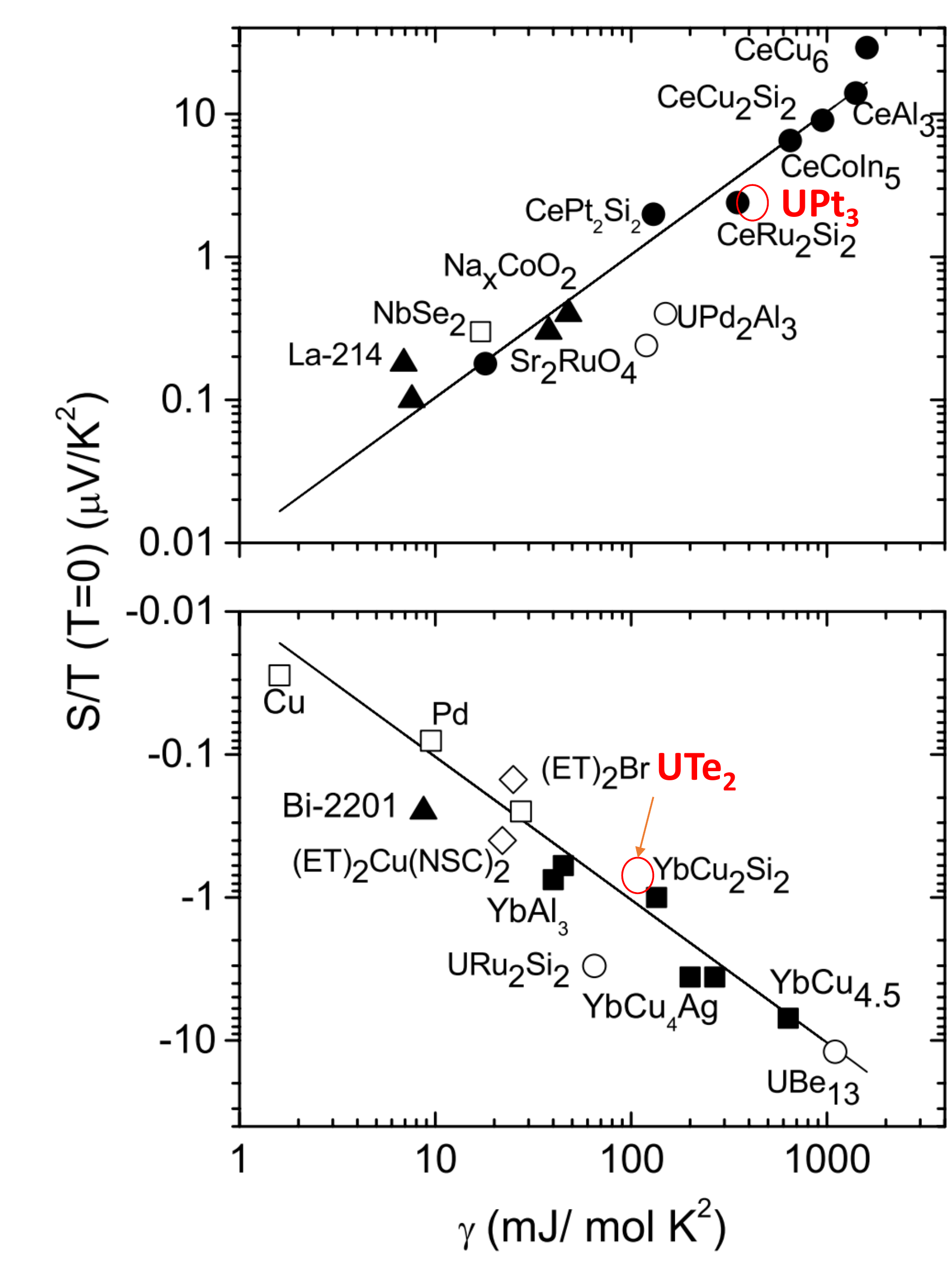}
 \caption{A correlation between the low-temperature slope of the Seebeck coefficient and the electronic specific heat is visible in dense metals (i.e. those with one carrier per formula unit).  The figure from ref. \cite{Behnia_2004} has been complemented with data on UPt$_3$ from ref.\cite{Zhu2009} and on UTe$_2$ from ref.\cite{Niu2020}. The plot shows that knowing the the Fermi energy of a metal,  one can predict (the rough magnitude of) the low-temperature Seebeck coefficient, which can be either positive (upper panel) or negative (lower panel).}
 \label{fig_Seebeck}
 \end{figure}

Remarkably, this simple formula is all one needs to predict the rough magnitude of the Seebeck coefficient at very low temperatures. In a wide variety of metals, experiments find a correlation between the magnitude of the electronic specific heat  and the low-temperature slope of the Seebeck coefficient  \cite{Behnia_2004} (See Fig. \ref{fig_Seebeck}). In these dense metals, the electronic specific heat is simply proportional to the inverse of the Fermi energy and this correlation indicates the validity of Eq. \ref{S1}  at sufficiently low temperature.
 
\begin{figure}[tbp]
 \centering
 \includegraphics[width=1\linewidth]{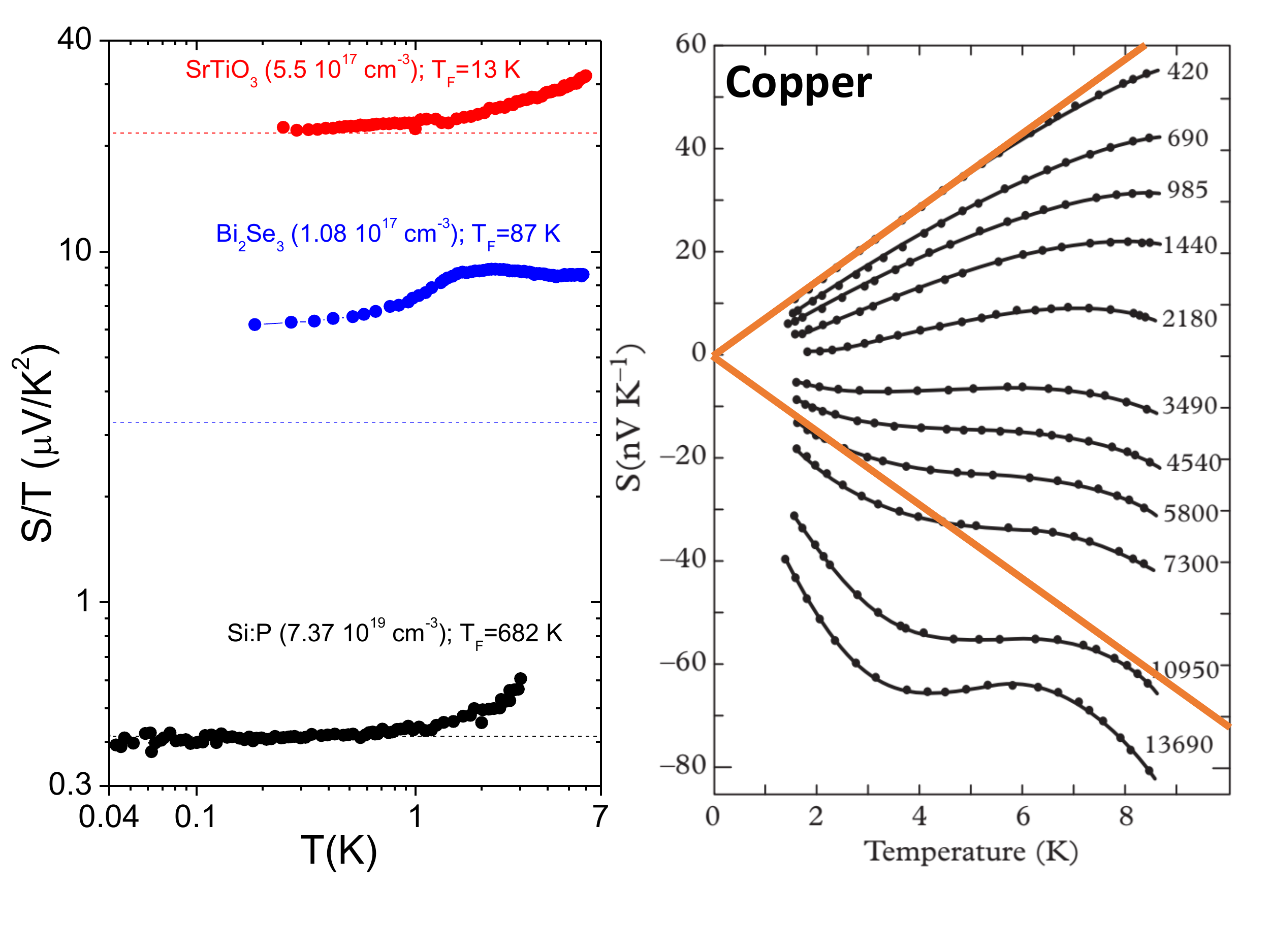}
 \caption{Left: The Seebeck coefficient in three dilute metals at sub-Kelvin temperatures. In the case of phosphorus-doped silicon, the sample has a carrier density above the metal-insulator transition \cite{Lohneysen}. The data for Bi$_2$Se$_3$ \cite{Fauque2013}  corresponds to a sample with anti-site doping, which makes it metallic with an observable Fermi surface. The data for SrTiO$_3$ corresponds to an oxygen-reduced sample with a detectable Fermi surface \cite{Lin2013}. In each case, the expected magnitude of the diffusive $S/T$ (assuming an energy-independent scattering time) is shown by horizontal dash lines. Right: Low-temperature Seebeck coefficient in copper for samples with different residual resistivities. The numbers give the ratio of room-temperature to residual resistivity \cite{Rumbo1976}. The solid orange lines show  the expected positive or negative diffusive $S$, given the Fermi temperature of copper.}
 \label{fig_dil}
 \end{figure}

 In dilute metals, where one mobile electron is shared by thousands of atoms, the electronic specific heat per moles is not normalized by the density of electrons and is no more simply the inverse of the Fermi energy. However, the latter can still be quantified by measuring (or calculating) the radius of the Fermi surface. The experimentally-measured  Seebeck coefficient divided by temperature, $S/T$ is close to what is expected in several heavily-doped semiconductors. Fig.\ref{fig_dil} shows three insulators, sufficiently doped to be on the metallic side of the metal-insulator transition where Eq. \ref{S1} is valid \cite{Lohneysen,Lin2013,Fauque2013}. This approach works even in a conventional metal such as YBa$_2$Cu$_3$O$_y$(p=0.11). There also, the amplitude of low-temperature $S/T$ agrees with the Fermi temperature quantified by the combination of the frequency of quantum oscillations and the effective mass of electrons \cite{Laliberte2011}.
 
 On the other hand, as seen in the right panel of the same figure, the  thermoelectric response in copper (much smaller because of the much larger Fermi energy) is extremely sensitive to the mean-free-path change. This indicates that  even at these low temperatures, the energy dependence of the scattering time matters. Nevertheless, the magnitude of the thermoelectric response remains in the expected range. 
 
\subsection{Two distinct versions of mobile entropy}
It is instructive to write the Landauer expression for thermal conductivity:

 \begin{equation}
 \kappa = A_d \frac{\pi^2}{3}\frac{k_B^2T}{h}k_F^{d-1}\ell
 \label{kappa}
\end{equation} 
 
 Let us compare first this equation to Eq. \ref{Land}. They obey the Widemman-Franz correlation between change and heat transport. In both cases, transmission scales with the mean-free-path. The  difference is on the identity of the transmitted quantity, the quantum of electric conductance $vs.$ the quantum of thermal conductance.
 
More instructive to our purpose is a comparison between Eq.\ref{kappa} and Eq.\ref{S2}. Both equations refer to the propagation of entropy in a medium. The difference resides in the way transmission occurs. Eq. \ref{S2} refers to a propagation exclusively driven by particle flow. The particles in question can accumulate entropy because of interactions between them. The mobility of this entropy will show itself only in the amplitude of the Seebeck coefficient (and not in the amplitude of thermal conductivity). In metals hosting heavy electrons,  thermal conductivity still depends on the mean-free-path  (and not the entropy) of their electrons. Their specific heat and Seebeck coefficient are both pushed up by the heaviness of the electrons (Fig. \ref{fig_Seebeck}).

 \subsection{Above the degeneracy temperature}
According to Eq. \ref{S1} (or equivalently Eq. \ref{S2}), the entropy of a degenerate mobile fermion is much smaller than the Boltzmann constant. In contrast, the Seebeck coefficient above the degeneracy temperature can be written as \cite{Collignon2020}:
\begin{equation}
S_{(\Lambda \ll \lambda_F)}= \pm \frac{k_B}{e}[C+\textrm{ln}\frac{\lambda_F^3}{\Lambda^3}]
 \label{S3}
\end{equation} 
Here $C$ is a constant, of the order of unity representing the degrees of freedom. This is a rearranged version of what is often called the Pisarenko equation\footnote{`Pisarenko' is a mysterious scientist working in the former Soviet Union presumably in the early 50s. His name has been frequently invoked to describe the variation of thermopower with carrier concentration in semiconductors (See  for example \cite{Gao2014}). Plausibly, this is a consequence of a few laconic sentences in an influential early book by Ioffe \cite{Ioffe1957}. Nevertheless, the 'Pisarenko equation' is already visible in a paper by Johnson and Lark-Horovitz as early as 1953 \cite{Johnson}. Discussing the thermoelectricity of extrinsic germanium, they write (see their equation 29): $S =\pm \frac{k_B}{e}[2-\textrm{ln}\frac{n\Lambda^3}{2}]$.

In Eq.\ref{S3} of the present paper, the carrier density is replaced by the Fermi wavelength, using $[3\pi^2n]^{1/3}=\frac{2\pi}{\lambda_F}$ in order to make visible the continuity across the degeneracy temperature.}. This equation has been used to understand the amplitude of the Seebeck coefficient in the so-called 'extrinsic' semiconductors, that is when the transport properties of the semi-conductor is controlled by the presence of extrinsic dopants. The entropy per electron appearing in this equation is intimately linked to the Sackur-Tetrode entropy \cite{Panos2015} of a classical Boltzmann gas.  Comparing Eq.\ref{S2} and Eq.\ref{S3} is a source of insight on fermionic entropy in quantum and classical contexts.

\subsection{Entropy, areas and volumes}
According to Eq. \ref{S3}, the entropy of a non-degenerate electron, on top of the constant due to equipartition, includes a term proportional to the logarithm of the ratio of two volumes. This is a consequence of particle indistinguishability, providing additional missing information about an electron identical to all others. 

Deep in the quantum regime, on the other hand, entropy per carrier is simply proportional to the ratio of two areas (Eq. \ref{S2}). The square of the Fermi wavelength quantifies the amount of missing information and the square of the thermal wavelength is the unit for this information. The most celebrated case of area-dependent entropy \cite{Elsert2010} is the Bekenstein-Hawking entropy of a black hole, proportional to the square of the event horizon  divided by the square of the Planck length \cite{bekenstein}. However, there are other cases of area-dependent entropy \cite{Srednicki1993,Elsert2010}.  \footnote{The Seebeck coefficient is an intensive quantity (proportional to the inverse of Fermi temperature). The area dependence of the entropy \textit{per carrier} does not impede the \textit{total} entropy of \textit{all} carriers to be extensive.}

Eq. \ref{S2} can be examined in the light  a viewpoint advocated by Jaynes \cite{Jaynes1957}, where the information theory provides the basis of statistical mechanics. The equation refers to a specific case  of thermodynamic entropy as missing information \cite{Beck2009}. The entanglement entropy of a degenerate fermion scales with its boundary area in the entire quantum system \cite{Elsert2010}. This is the square of the Fermi wavelength normalized by the square of the thermal wavelength. The latter quantifies the uncertainty in the position of a particle with a thermal momentum \cite{Silvera}.
\subsection{Thermoelectricty of insulators}
In a band insulator, carriers of heat and charge are electrons or holes, which are thermally activated across the band gap, $E_G$. Thus, a carrier of charge $e$ has a thermal energy equal to the half of the band gap and the Peltier  Coefficient is  $\Pi=E_G/2e$. The Kelvin relation then implies that $S=\pm\frac{k_B}{e}\frac{E_G}{2k_BT}$ \cite{MacDonald,Behnia2015b,Machida2016}. 

A Seebeck coefficient steadily enhancing with cooling has indeed been experimentally detected in several insulators \cite{Geballe1954,Geballe1955,Mortensen,Machida2016}. A theory of polarization-driven thermoelectric response has been recently put forward \cite{onishi2021theory}. A major challenge in this field of investigation is the unavoidable presence of a large 'phonon drag' term, on top of the diffusive response \cite{Herring}. Even in the case of elemntal insulators, such as germanium \cite{Geballe1954} or silicon \cite{Geballe1955}, this complicates the analysis.

\section{The Nernst coefficient}
The flow of heat and charge is expressed by two equations linking three tensors and four vectors:
\begin{equation}
\begin{split}
    \mathbf{J_{e}}=\overline{\sigma}. \mathbf{E} - \overline{\alpha}.\mathbf{\nabla T} \\
    \mathbf{J_{q}}= T \overline{\alpha} . \mathbf{E} - \overline{\kappa}.\mathbf{\nabla T}
\end{split}
\end{equation}
Here $\mathbf{J_{e}}$ is the charge  density current and  $\mathbf{J_{q}}$ is the heat density current.  $\mathbf{E}$ is the electric field and  $\mathbf{\nabla T}$ is the thermal gradient. The three conductivity tensors are electric, $\overline{\sigma}$, thermoelectric, $\overline{\alpha}$,  and thermal, $\overline{\kappa}$, which acquire an off-diagonal component in presence of magnetic field. The Nernst effect refers to the off-diagonal component of  $\overline{\alpha}$. 
  \begin{figure}[tbp]
 \includegraphics[width=0.9\linewidth]{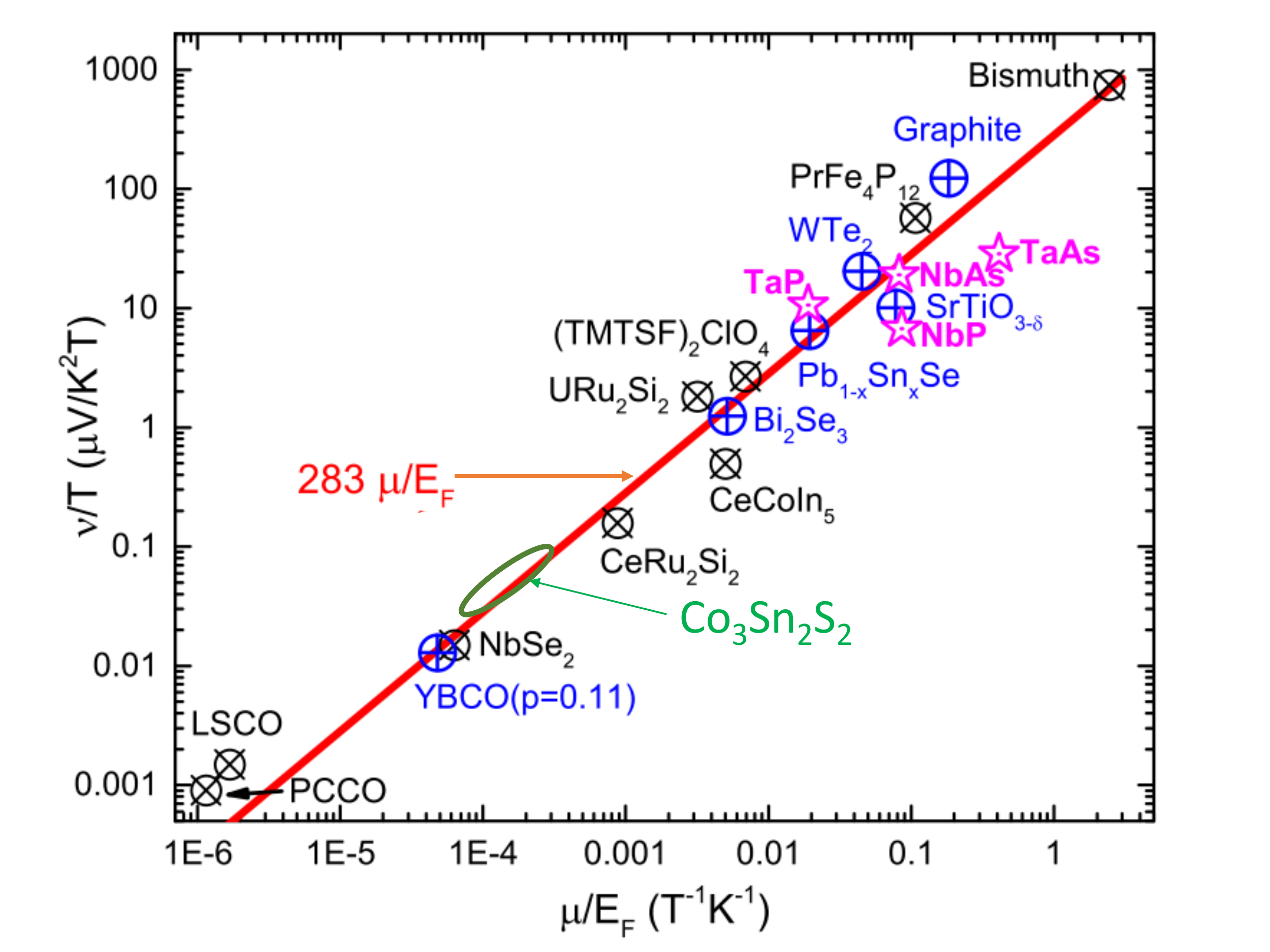}
 \caption{The amplitude of the slope of the low-temperature Nernst coefficient in a variety of metals vs. the ratio of mobility to Fermi energy. The data in ref. \cite{Behnia2016} is complemented by one topological and magnetically ordered semimetal (Co$_2$Sn$_2$S$_2$\cite{Ding2019}) as well as four non-magnetic ones (TaAs, TaP, NbAs, and NbP \cite{Xu2021}). The plot shows that knowing the mobility and the Fermi energy of a metal, one can predict the rough magnitude of the low-temperature Nernst coefficient.}
 \label{fig_Nernst}
 \end{figure}
The experimental procedure is to apply a longitudinal thermal gradient and to measure the transverse electric field.  The ratio of these two is the Nernst signal ($S_{xy}=\frac{E_y}{\nabla_xT}$). The Nernst coefficient, $\nu$, is this signal divided by the magnetic field. Combining $S_{xy}$ with components of $\overline{\sigma}$, the conductivity tensor, one can extract $\alpha_{xy}$. In absence of stabilized denomination, the latter will be referred as 'Nernst conductivity', in the following. Its fundamental units are $\frac{k_{B}e}{h}= 2.2$ nA/K. We will focus on its expression in two dimensions.
 
\subsection{The quasi-particle Nernst signal}
 According to the Mott relation, the thermoelectric response is set by the energy derivative of the conductivity at the Fermi level \cite{Mott1936}. This links $\alpha_{xy}$  to $\sigma_{xy}$:
 
 \begin{equation}
 \alpha_{xy} = \frac{\pi^2}{3} \frac{k_B}{e}k_B T\frac{\partial \sigma_{xy}}{\partial \epsilon}\mid_{\epsilon=\epsilon_f}
 \label{Hall}
\end{equation}

 The latter can be expressed using the magnetic length  derived from the magnetic field $B$ and fundamental constants: $\ell_B=\sqrt\frac{\hbar}{eB}$. In two dimensions, one has : 

\begin{equation}
 \sigma_{xy} \approx \frac{e^2}{h} \frac{\ell^2}{\ell_B^2}
 \label{Hall}
\end{equation} 

What is the amplitude of the change in $\sigma_{xy}$ induced by an infinitesimal shift in the position of the chemical potential?  Assuming a smooth energy dependence for the Fermi wavevector and the mean-free-path near the Fermi energy, the Mott relation will lead us to the following expression \cite{Behnia2016}:  
\begin{equation}
\alpha_{xy} \approx \frac{k_{B}e}{h} \frac{\lambda_{F}^2}{\Lambda^2}\frac{\ell^2}{\ell_B^2}
\label{N1}
\end{equation}

According to Eq. \ref{N1}, the amplitude of $\alpha_{xy}$ is set by the ratio of the Boltzmann constant and the quantum of magnetic flux, ($\frac{k_{B}e}{h}$). This natural scale is multiplied by a combination of four length  scales. The magnetic length, $\ell_B$ depends only on the magnetic field, but the three other are material-dependent. Long mean-free-path, short de Broglie wavelength and long Fermi wavelength enhance $\alpha_{xy}$. The record of the largest Nernst signal generated by degenerate electrons is still held by bismuth \cite{Heremans1976,Behnia2007}, the solid in which Nernst and Ettingshausen first discovered this effect. The fundamental reason behind this fact is the lightness of electrons and the diluteness of electron fluid in bismuth. Amazingly, this understanding was not explicitly formulated before 2007 \cite{Behnia2007,Behnia2009}. 

Eq. \ref{N1} points to a correlation between the amplitude of $\alpha_{xy}$ and material-dependent properties in two dimensions. Now,  what is experimentally measured is the Nernst signal, S$_{xy}$,  equal to $\alpha_{xy}$ multiplied by electrical resistivity. According to Eq. \ref{N1} (and independent of dimensionality) the slope of the  coefficient should be proportional to the ratio of mobility to the Fermi energy.  Experimental data  gathered by numerous experiments on metals with different mobilities and Fermi energies have confirmed this correlation between the magnitude of the Nernst coefficient ($\nu=S_{xy}/B$) at low temperature and the ratio of mobility to Fermi energy (See Fig.\ref{fig_Nernst}) \cite{Behnia2009,Behnia2016}.

Callen identified the Seebeck coefficient as a measure of entropy carried by a mobile carrier of charge in absence of thermal gradient. Following the same approach, the Nernst conductivity (i.e. the transverse thermoelectric coefficient, $\alpha_{xy}$) can be defined as a measure of entropy carried by a mobile carrier of magnetic flux (and in absence of thermal gradient). 

Let us now apply this approach to two other cases. In the first case, the magnetic flux is `fictitious' and in the second case the carrier is not an electronic quasi-particle.  
\subsection{The anomalous Nernst effect and the Berry Curvature} 
Eq. \ref{Hall} is the semi-classical expression for the `ordinary' Hall effect. In ferromagnets, there is an additional component to the Hall response, dubbed `anomalous' or 'spontaneous' \cite{ADAMS}, which is finite even at zero applied magnetic field. The origin of this latter component  has been traced to the Berry curvature of the electron wave-function \cite{Nagaosa2010,Xiao2010,Sinitsyn}. 

The ordinary Hall effect arises because the magnetic field deflects the trajectory of electrons. 
The emergence of Berry curvature leads to a `fictitious' magnetic field giving rise to the anomalous Hall conductivity \cite{Nagaosa2010,Xiao2010}:

\begin{equation}\label{AHC}
\sigma^{A}_{xy}=-\frac{e^{2}}{\hbar}\int_{BZ}\frac{d^{D}k}{(2\pi)^{D}}f(k)\Omega_B(k)\approx -\frac{e^2}{\hbar} <\frac{\Omega_B}{\lambda_F^2}>
\end{equation}

Here, $\Omega_B$ is the Berry curvature, defined as the curl of Berry connection. The latter is defined as the projection a Bloch function over its derivative in the momentum space \cite{Gosalbez2015,Price2014}. $\Omega_B$ has the dimensions of an area, and for each band, the periodicity of the reciprocal lattice. The brackets in this equation schematize an averaging process over the whole Fermi sea. The anomalous Hall conductivity is the quantum of conductance multiplied by the ratio of two areas: The square of the fictitious magnetic length (i.e. the Berry curvature) and the square of the Fermi wavelength.

The thermoelectric counterpart of this phenomenon is the anomalous Nernst effect, which refers to a finite  $\alpha_{xy}$ even in absence of magnetic field. Its expression can be simplified to \cite{Ding2019}: 
\begin{equation}\label{ANC}
\alpha^{A}_{xy} \approx \pm \frac{ek_B}{h} <\frac{\Omega_B}{\Lambda^2}>
\end{equation}

Expressing these transport coefficients as the product of fundamental units and material-dependent length scale leads to two insights about experimental facts.

The first is about the way disorder affects ordinary and anomalous Nernst responses. According to Eq.\ref{AHC}, $\alpha^{A}_{xy}$ does not depend on the mean-free-path, in contrast to the ordinary $\alpha_{xy}$ (See Eq. \ref{N1}). Now, what is experimentally measured is the Nernst signal, which is $\alpha_{xy}$ divided by resistivity. Therefore, in a given solid, increasing the mean-free-path should have opposite consequences for the two components of the Nernst signal. It should deplete the anomalous $S_{xy}$, while enhancing the ordinary  $S_{xy}$. This is indeed what has been observed in Co$_3$Sn$_2$S$_2$ \cite{Ding2019}.

\begin{figure}
 \includegraphics[width=0.75\linewidth]{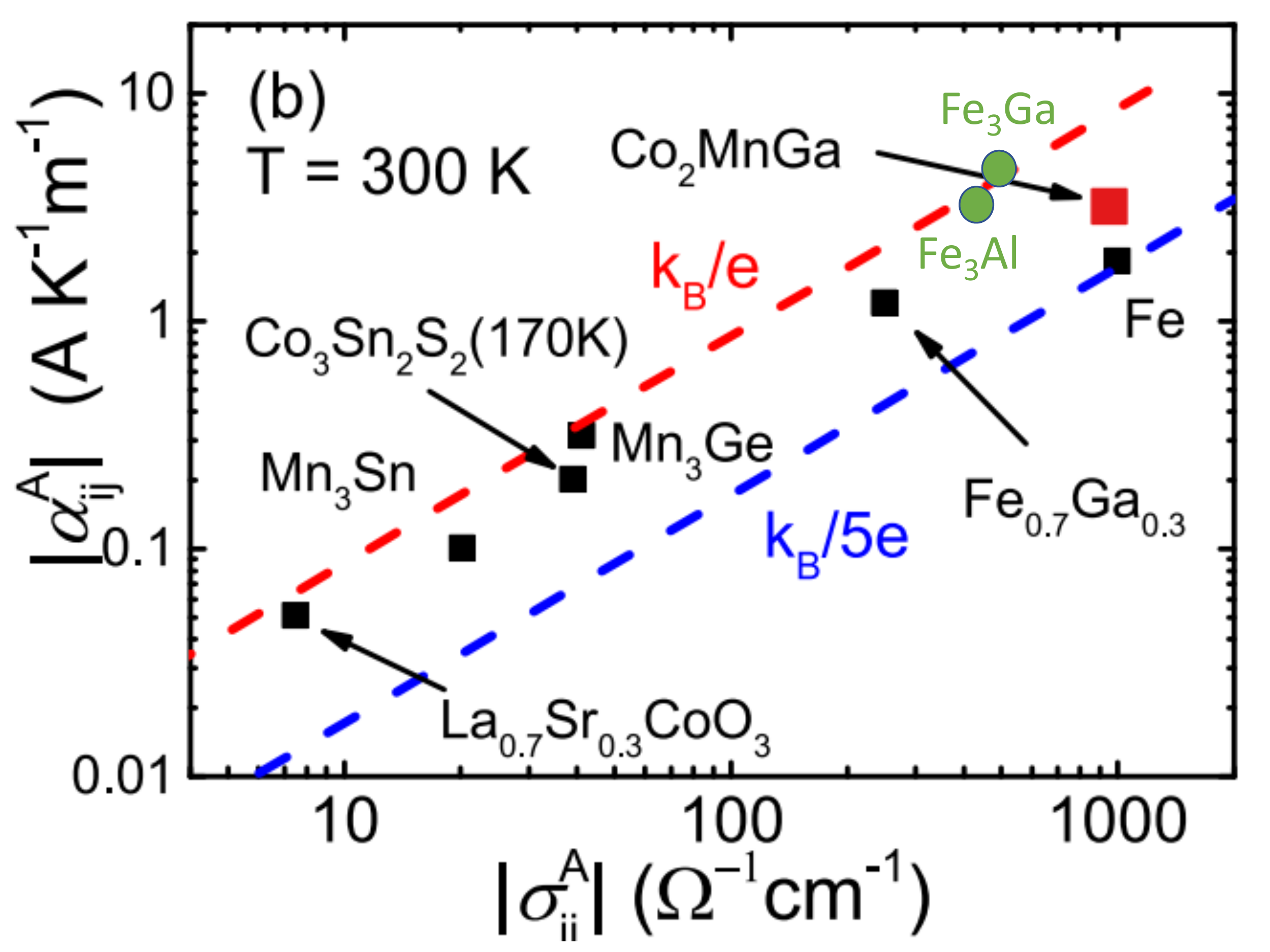}
 \caption{The anomalous Nernst conductivity at room temperature $vs.$ the anomalous Hall conductivity in a variety of topological magnets. A correlation is visible. Data from ref.\cite{Liangcai} is complemented by data on Fe$_3$Ga and Fe$_3$Al from ref.\cite{Sakai2020}.}
 \label{fig_ANE}
 \end{figure}

The second feature concerns an expected correlation between the anomalous Nernst conductivity and the anomalous Hall conductivity. Comparing Eq.\ref{AHC} and Eq. \ref{ANC}, one would expect  the ratio of the two anomalous conductivities to scale  with the ratio of the square of the two wavelengths.  $\frac{\alpha^{A}_{xy}}{\sigma^{A}_{xy}}$ should vanish at low temperature, and should tend towards $\frac{k_B}{e}$ at high temperature, because $\lambda_F$ will eventually approach $\Lambda$. Such a trend is experimentally observed across topological magnets \cite{Liangcai,Asaba} (See Fig.\ref{fig_ANE}).  Note, however, that in all these cases, the Fermi temperature is still much larger than room temperature and the Fermi wavelength longer than the de Broglie thermal length.
\subsection{The Nernst effect and superconductivity}
Understanding the origin of the Nernst effect in various solids became a quest following the observation of an enigmatic Nernst signal in underdoped cuprates  and enthusiastically attributed to ``vortex-like excitations above T$_c$''\cite{Xu2000} at the turn of this century. It was then  believed that `` the Nernst signal is generally much larger in ferromagnets and superconductors than in nonmagnetic normal metals.''\cite{Wang2006}. In a remarkable episode of collective amnesia, the large amplitude of the Nernst effect in bismuth, which made possible its discovery with the technology of nineteenth century and decades before the discovery of superconductivity, was forgotten. 

Nevertheless, Nernst effect proved to be a sensitive probe of the  superconducting fluctuations  above the critical temperature of \textit{any} superconductor. Soon, Ussishkin, Sondhi and Huse \cite{Ussishkin} calculated that the Gaussian fluctuations of the superconducting order parameter leads to a contribution to the Nernst conductivity, which was  expressed as: 
\begin{equation}
\alpha^{SC}_{xy} \approx \frac{k_Be}{h} \frac{\xi^2}{\ell_B^2}
\label {USH}
\end{equation}
Here, $\xi$ is the superconducting coherence length. What produces the Nernst response are short-lived Cooper pairs above T$_c$, capable of carrying magnetic flux and entropy. 

Comparing Eq.\ref{USH} with Eq. \ref{N1}, one sees that a signal of superconducting origin in the normal state will dominate in dirty and dense superconductors. Indeed, experiments on such a superconductor found an excellent agreement between theory and experiment \cite{Pourret2006}.  In thin films of Nb$_{0.85}$Si$_{0.15}$, a Nernst signal caused by short-lived Cooper pairs was detectable up to temperatures exceeding the critical temperature by a factor of 30 \cite{Pourret2006}. Up to twice T$_c$, the magnitude of this signal was in excellent agreement with Eq.\ref{USH}. Subsequent analysis showed that the data in the normal state is governed by a single length scale constructed by a combination of the superconducting correlation length and the magnetic length \cite{Pourret2007}. 

\begin{figure}
 \includegraphics[width=0.95\linewidth]{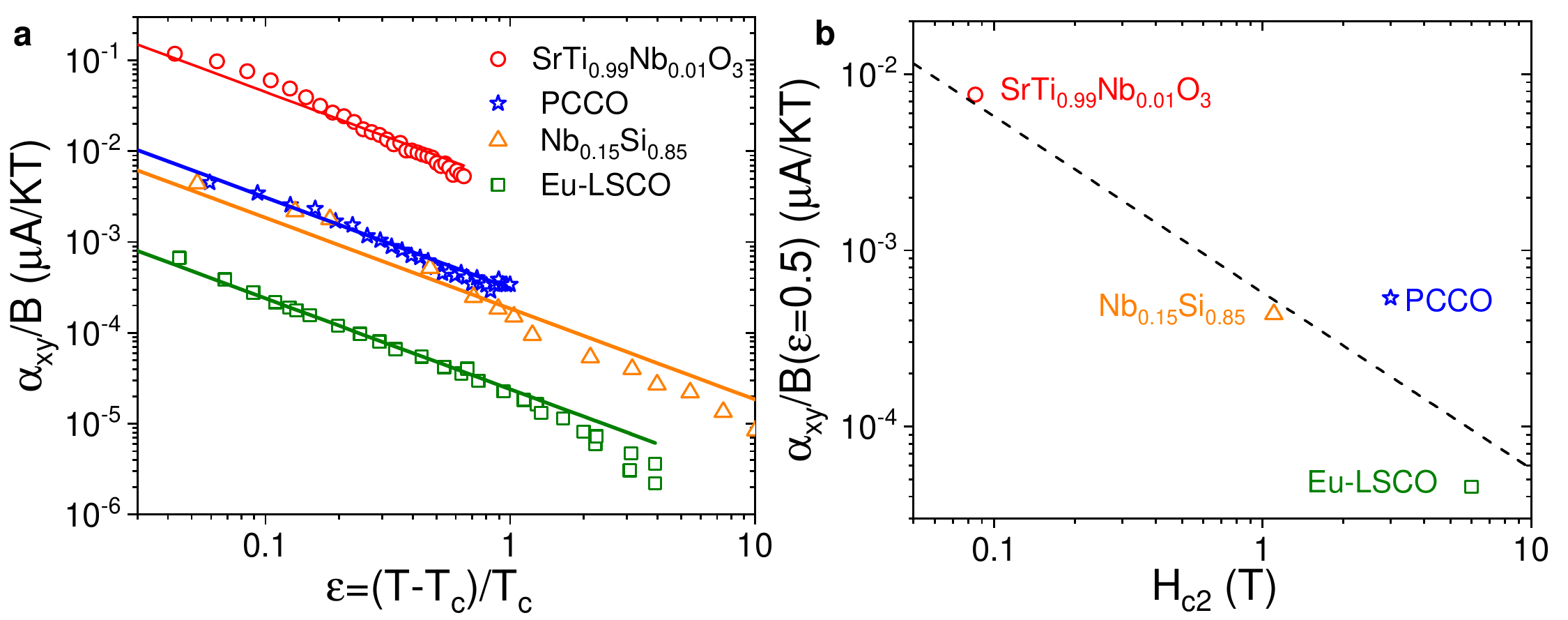}
 \caption{a) The temperature dependence of the Nernst conductivity generated by fluctuating Cooper pairs in four different superconductors \cite{Pourret2006,Chang2012,Tafti2014,Rischau2021} as a function of reduced temperature (solid lines represent the theoretically expected amplitudes. b) The amplitude of the fluctuating signal in the normal state of these superconductors at T=1.5T$_c$ as a function of their upper critical field. The latter is proportional to the inverse of the square of the superconducting coherence length.}
 \label{fig-SC}
 \end{figure}

The original theory was followed by other  more elaborate theoretical efforts \cite{Serbyn2009,Michaeli2009}. They confirmed the validity of the original approach \cite{Ussishkin}. On the experimental side, extensive studies of the Nernst effect in cuprates \cite{kakonovic2009,Chang2012,Tafti2014,Cyr2018}  demonstrated that the fluctuating Nernst signal in those materials can also be described in the Gaussian picture (See Fig. \ref{fig-SC}). 
 
What remains most enigmatic about the superconducting Nernst signal is its amplitude below T$_c$ \cite{Rischau2021}. It is understood that the fluctuating signal above the critical temperature depends on a unique material dependent length scale, which is $\xi$. On the other hand, below T$_c$. the observation that the amplitude of the signal in the vortex liquid regime is roughly similar among different superconductors remains a mystery. The extracted irreducible entropy of a mobile superconducting vortex is found to be of the order of a Boltzmann constant per layer in numerous superconductors \cite{Rischau2021}. Understanding this feature is a subject matter of ongoing research. 

Given that the study of the Nernst effect in this century kick-started with the assumption that mobile superconducting vortices are the major source of a Nernst signal, it is ironic that our present understanding of the vortex Nernst signal lags behind what we know about the other sources of the Nernst signal. 

\section{Acknowledgements} I am grateful to Jacques Flouquet and Didier Jaccard who initiated me to thermoelectricity. I would like to thank Louis Taillefer, Michel Ribault, Herv\'e Aubin, St\'ephane Belin, Cigdem Capan, Saco Nakamae, Cyril Proust, Nigel Hussey, Romain Bel, Hao Jin, Alexandre Pourret, Yakov Kopelevich, Luis Balicas, Koichi Izawa, Aritra Banerjee, Huan Yang, Zengwei Zhu, Elena Hassinger, Yuki Fuseya, Xiao Lin,  Aur\'elie Collaudin, Adrien Gourgout, Lisa Buchauer, Yuke Li, Yo Machida, Valentina Martelli, Cl\'ement Collignon, Alexandre Jaoui, Aharon Kapitulnik, Harold Hwang, Xiaokang Li, Liangcai Xu, Binghai Yan and Felix Spathelf, who were my fellow travellers in exploring thermoelectric and thermal properties of numerous solids. A special thanks goes to  Beno\^it Fauqu\'e for his stimulating presence and his role in driving our multiple common projects since 2008.
\bibliographystyle{crunsrt}
\bibliography{biblio}
\end{document}